# Theoretical Chemistry Course for Students in Chemistry

Qingyong Meng*

Department of Chemistry, Northwestern Polytechnical University, West Youyi Road 127, 710072 Xi'an, China

**ABSTRACT**
In this work, the teaching content of a theoretical-chemistry (TC) course is reformed, establishing a theoretical contents from micro- to macro-system, and comprehensively introducing the theory of chemical reaction to undergraduate students in chemistry. In order to develop such TC course based on the general physical-chemistry course, we focus on the "last-mile" problem between the physics and chemistry courses to train the critical thinking of undergraduate students in chemistry. To clearly show this, a reduction scheme of polymer molecular dynamics was discussed as an example, which shows a different theoretical content in polymer chemistry. Moreover, we propose a series of experiences and dependent measures that can provide information regarding students' levels of knowledge and understanding. This assessment quiz was designed to test students on the fundamental concepts and applications of TC, such as dynamics, statistical ensemble, kinetics, and so on. From the actual teaching for 36 students, it was found that these students performed significantly improvement from the present TC content. Further analysis of each individual question revealed that approximately two-third of the students learn new knowledge. Although the present TC course might be considered to be a certain degree of difficulty for chemists, these analyses show that students can effectively accept these complicated concepts.

## INTRODUCTION

Over the past decade, calculations on modern computer have become an invaluable tool for chemistry studies and hence computational chemistry has been an important research field.[1-11] The computational chemistry not only tests new theoretical ideas on systems far too complex for manual calculation, but also approximates laboratory experiments to lead a microscopic insight into realistic experiments and provide virtual access to extreme conditions of the real world.[10,11] The power of computational chemistry derives largely from the fact that it occupies a unique position between the regimes of theoretical physics and laboratory experiment for chemical reactions. Therefore, it is a prerequisite of a modern chemist to learn enough theories and computational technologies in chemistry[12-18] as well as the modern engineering technology such as artificial intelligence and quantum computing.[19,20]

However, the general physical chemistry course[21-23] can not yet provide enough theoretical chemistry for the modern chemist, in particular theoretician, which implies further development of the teaching content for the theories in chemistry. In this work, a teaching scheme is proposed where we rebuild the teaching content of theoretical physics[12-18] for chemist. This is so-called theoretical chemistry (TC) course in this work. We believe that a TC course should include the following content that will be discussed in detail. First of all, the TC course gives the basic theory of dynamics, including classical and non-relativistic quantum theory, deriving the elemental movement of a molecule such as vibration, rotation, and scattering in both classical and quantum viewpoints. Second, as a bridge between macroscopic and microscopic system, the TC course has to also focus on the statistical mechanics for not only equilibrium thermodynamics but also molecular kinetics. Having given the above two parts, the electronic structure theory must be given and then teach molecular Hamiltonian operator under or beyond Born-Oppenheimer approximation.[10,24] In many institutions, the electronic structure theory together with Born-Oppenheimer approximation are given in a separated course that is often called quantum chemistry (QC) or electronic structure theory. In parallel with the QC course, as the third part of the present TC course, construction methods of the Hamiltonian operator has to be included. Then, on the basis of the Hamiltonian operator, classical and quantum molecular dynamics[11,25] must be given as the fourth and the last part, where the numerical technology of solving time-dependent Schrödinger equation (TDSE) for elementary reactions should also be discussed in detail. In this part, especially in solving TDSE, the main goal of teaching is chemistry dynamics and kinetics, such as reactive probability, reaction rate constant, and mechanisms. Besides of the above content, in actual teaching, an interesting question is whether molecules know how to solve these equations of motion (EOMs) or not. The answer is clearly no. To clearly explain this question from science, most of descriptions for the theoretical framework build on the basis of principle of least action and/or path integral,[26,27] as well as the statistical laws of probability.[16-18,28-30] We believe that the present TC course is an advanced or special topics course, offering a learning opportunity for undergraduates, while this course provides a outline of chemistry theory making it also be suitable as a graduate course.

Obviously, all of above parts mainly focus on dynamics and kinetics of chemical reactions, which need lots of knowledge of mathematics and computational technology, such as linear algebra, integral equations, differential equations, program language, and so on. Thus, the TC course provides a bridge connecting mathematics, physics, and chemistry. And hence, students in the TC course must be able to use various mathematical tools to abstractly explain and/or demonstrate concepts in this course, which implies that the TC course is helpful to train the capability of critical and abstractly thinking. Nevertheless, such capability is often measured by an individuals' ability to describe theoretical-chemical concepts in mathematics. To semi-quantitatively investigate this capability of students and obtain learning feedback, we further propose a series of dependent measures that can provide information regarding students' levels.

The rest of this paper is organized as follows. In Section II, we will describe the currently available theoretical courses for chemist. Section III presents the theoretical chemistry in education and shows teaching content of the present TC course. In Section IV we give methodology for evaluating the students' levels and discuss the evaluation results. Finally, Section V concludes with a summary.

**CURRENTLY AVAILABLE THEORETICAL-CHEMISTRY COURSES**

Chemistry is an old but fast developing field with so many branches to learn for a undergraduate student,[23] which implies difficulty to add a course in ever-more-crowded undergraduate curriculum. As is well known, as a field of sciences, chemistry is generally divided into experimental, theoretical, and computational fields.[21-23] With the fast development in the interdisciplinary field of chemistry with material, biology, information technology, electrical engineering, *etc.*, the chemistry theory has already been largely expanded to a very depth level,[23] which contains not only the traditional physical chemistry but also many branches of theoretical physics.[12-18]

Currently, in the undergraduate courses, contents of chemistry theory include (statistical) thermodynamics, chemistry dynamics and kinetics, structure of molecule and material, primitive introduction of the electronic structure theory, and so on.[21,22] These teaching contents are included in several chemistry courses, say general chemistry, physical chemistry, statistical thermodynamics, chemistry dynamics and kinetics, structure of the matter, and other specialized courses. In addition, the fresh students in chemistry also have to finish a two-semesters course of college physics,[12,31,32] which mainly contains classical mechanics, thermodynamics, electrodynamics, quantum mechanics, *etc.* Generally, the currently available physical-chemical courses as well as college physics give almost all necessary content for students in chemistry,[12,31,32] even leading to repeated learning time for some important point. For example, thermodynamics for equilibrium has been deeply discussed in college physics, inorganic chemistry and/or college chemistry, and then has been emphasized in physical chemistry.[12,21,22,31,32]

Nevertheless, a large number of important theories for modern theoretical chemistry, such as dynamics, statistical ensemble, quantum molecular dynamics, numerical algorithms, *etc.* have been discarded, while currently developed engineering technology have not been introduced in the courses. These features of the currently available physical-chemistry courses for the undergraduate students can no longer meet the existing requirements of the developed engineering technologies and interdisciplinary subjects. For example, electron spin and Larmor precession are the keys to understand nuclear magnetic resonance. In analytical chemistry course,[33] spin is generally erroneously drawn as a behavior similar to earth's rotation, while in college physics course Larmor precession is not introduced. This is a deviation in the knowledge structure of students. In fact, these contents only need to extend college physics slightly. This is a so-called "last-mile" problem between physics and chemistry courses. In this work, we shall consider this problem and try to build a reasonable teaching content to resolve this problem from the viewpoint of dynamics and kinetics.[34-39] This content, that is the TC course, can be taught just after the general physical chemistry as well as college physics, which also provide sufficient knowledge reserves for the present TC course.

**THEORETICAL CHEMISTRY IN UNDERGRADUATE EDUCATION**

In this section we propose a scheme of the TC course, which needs at least 40 academic hours, including about 8 academic hours for experiences and examination. Given in Tables 1 and 2 are main teaching contents, while Table 3 gives examples of experiences. Many of these content is taught through a lecture-based course, while the students are asked to take notes by themselves. Now, let us show main teaching content in the present TC course, and then show an example of the experiences.

**Table 1. Comparison of the contents in the first part of theoretical chemistry with the other theory courses, say college physics and physical chemistry. In this part, classical and quantum dynamics are included, together with the Born-Oppenheimer approximation. They need about 18 academic hours. The first column gives the modules of this course. The second column shows content number of the teaching contents. The third and fourth columns compare the contents of theoretical chemistry with those of college physics and physical chemistry. The rightmost column gives a suggested set of academic hours for every points.**

| Modules | No. | Contents in Detail | | Academic Hours |
|---|---|---|---|---|
| | | Theoretical Chemistry[a] | College Physics and/or Physical Chemistry[b] | |
| Classical Dynamics | | | | |
| | I | principle of least action | | 1 |
| | II | Lagrange and Hamilton EOMs | | 1 |
| | III | symmetry and conservation laws | | 1.5 |
| | IV | derivation of Newton's Laws | give Newton's Laws | 1 |
| | V | vibration, rotation, scattering | force, vibration, rotation | 2 |
| | VI | why needs quantum mechanics | | 0.5 |
| | | | | In total: 7 |
| Quantum Dynamics | | | | |
| | VII | measurement a micro-particle and path | | 1 |
| | VIII | particles don't need to first solve equation | | 1 |
| | IX | derivation of Schrödinger equation | Schrödinger equation | 1 |
| | X | vibration mode | vibration | 1.5 |
| | XI | rotation and spin | rotation | 1.5 |
| | XII | quantum scattering | | 1 |
| | XIII | quantum molecular dynamics | | 1 |
| | | | | In total: 8 |
| Born-Oppenheimer | | | | |
| | XIV | separation of electron and nuclear | Born-Oppenheimer | 1 |
| | XV | coupling between two electronic states | | 1 |
| | XVI | example: photochemistry mechanism | | 1 |
| | | | | In total: 3 |

[a]The proposed teaching contents in the present TC course. [b]See also references 12,21-23,31,32.

**Table 2. Same as Table 1, except for the second part. In this part, statistical theory for equilibrium and motion is included, together with the kinetics of reaction network. They need about 13 academic hours.**

| Modules | No. | Contents in Detail | | Academic Hours |
|---|---|---|---|---|
| | | Theoretical Chemistry[a] | College Physics and/or Physical Chemistry[b] | |
| Statistical Ensemble | | | | |
| | XVII | canonical ensemble theory | | 1 |
| | XVIII | quantum ensemble and wave function | | 1 |
| | XIX | statistical thermodynamics from ensemble | thermodynamics | 1.5 |
| | XX | partition function | partition function | 1.5 |
| | XXI | gas kinetics and energy distribution | gas kinetics | 1 |
| | | | | In total: 6 |
| Dynamics and Kinetics | | | | |
| | XXII | reactive scattering and intersection | | 1 |
| | XXIII | time-independent dynamics method | | 1.5 |
| | XXIV | time-dependent dynamics method | | 1.5 |
| | XXV | mechanism reduction | | 1 |
| | XXVI | dimensional reduction | | 1 |
| | XXVII | dynamics of reaction network | | 1 |
| | | | | In total: 7 |

[a]The proposed teaching contents in the present TC course. [b]See also references 12,21-23,31,32.

**Table 3. Examples of the experiences in the theoretical chemistry course, where several simple reactions are discussed by various methods. The experiences need about eight academic hours, while the examination needs one academic hour.**

| Modules | No. | Experiences | | Academic Hours |
|---|---|---|---|---|
| | | Reaction[a] | Question Overview | |
| Classical Dynamics | | | | |
| | i | $H + H_2/CH_4$ | do force analysis to see its failure for reaction | 0.5 |
| | ii | | Find its Lagrange and Hamilton equations | 0.5 |
| | | | | In total: 1 |
| Quantum Dynamics | | | | |
| | iii | $H + H_2/CH_4$ | find Schrödinger equation | 0.5 |
| | iv | | how about electronic motion | 0.5 |
| | v | | quantum scattering of molecules | 1 |
| | | | | In total: 2 |
| Born-Oppengeimer | | | | |
| | vi | $H + H_2$ | comparing electron with nuclear | 0.5 |
| | vii | | conical intersection of the $H + HD$ system | 0.5 |
| | | | | In toal: 1 |
| Statistical Ensemble | | | | |
| | viii | $H + H_2/CH_4$ | partition function from *ab inition* calculations | 0.5 |
| | ix | | Eyring equation for rate coefficient | 0.5 |
| | | | | In toal: 1 |
| Dynamics and Kinetics | | | | |
| | x | $H + H_2/CH_4$ | Quantum scattering of the $H + HD$ reaction | 0.5 |
| | xi | | solution of time-dependent Schrödinger equation | 0.5 |
| | xii | | obtain rate coefficient from reactive probability | 1 |
| | xiii | $H_2 + O_2$ | analysis of the reaction network in combustion of $H_2$ | 1 |
| | | | | In total: 3 |
| Examination | | | | In total: 1 |

[a]The reaction for understanding the question.

### Classical and Quantum Molecular dynamics

As is well known, classical molecular dynamics composes a large research field of the computational chemistry and molecular simulation,[8-11] implying classical mechanics plays an importance role in theoretical chemistry. Moreover, classical mechanics is an important gateway to theoretical physics, giving almost all of important concepts, such as principle of least action, Lagrangian, Hamiltonian, EOM, *etc*. Hence, almost all theoretical tutorials start with classical dynamics as given in Table 1. As the beginning of the present course, we discuss principle of least action after very fast showing concepts of coordinates and momentum. Then, the variational principle is discussed and then derives the Lagrange and Hamilton EOMs. Next, from the homogeneous and isotropic features of space as well the homogeneous feature of time, we derive the conservation laws of energy, momentum, and angle momentum and then show Nöther's theorem. From Galileo's relativity principle and the symmetry, we further derive celebrated Newton's laws, that is Newton EOM, that were acquaintances by students. In parallel with these derivations, concept of force is derived from EOM and then on the basis of the Lagrange, Hamilton, and Newton EOMs, typical motion in molecule, such as vibration, rotation, and scattering are considered. Finally, as a supplement material for teaching, classical relativistic mechanics is introduced form the above classical mechanics with the help of the Lorentz's relativity principle, and then classical theory of electromagnetic field and Coulomb's law are naturally introduced. All of the above contents on classical dynamics can be taught within roughly seven academic hours.

Next, from the fact that the molecule cannot know how to solve classical EOM to obtain a unique trajectory, we induce students to understand the failure of classical dynamics for molecular motions. Having shown this failure, we have to introduce quantum path and hence the concept of path integral to naturally resolve this failure. Then we can turn to non-relativistic quantum mechanics, which will take about eight academic hours for teaching. As is well known, chemical reaction is essentially a quantum process. The undergraduate students in chemistry must comprehensively and in-depth study the basis of quantum mechanics as an important theoretical foundation for understanding chemical reactions. In the present TC course, the reaching content for quantum mechanics mainly focus on the nuclear motions,[34-39] as introduction of the electronic-structure theory is often a part of physical chemistry.

In the present TC course, since the quantum mechanics of nuclear motions may be not conventional for chemist, we have to discuss it with many examples. To do this, after the concept introduction of path integral, we derive the Schrödinger equation of free particle to show EOM in quantum mechanics. By this equation, we first derive the eigen-problem of vibration of diatomic molecule and rotation of rigid body. Furthermore, taking $H_2^+$ and $H + H_2$ as examples we show the outline for solving the Schrödinger equation of steady and continuous states. Then we discussed quantum scattering of two-body system and molecule-surface system, which are simple models of the gas phase and heterogeneous reactions, respectively. Resolving Schrödinger equation of the $H + H_2$ reaction step-by-step by pseudo-code, we introduce the concept of quantum molecular dynamics, including potential energy surface (PES), reactive probability, mode-specific mechanism, and so on. Comparing the quantum molecular dynamics with the macroscopic phenomena (say temperature, rate constant, *etc.*) which is known well by undergraduate students, we naturally enter the bridge between microscopic and macroscopic motions, that is statistical mechanics, as an important module of the next part.

Standard Model in Chemistry

    Having shown the classical and quantum molecular dynamics, we further give statistical ensemble for chemical reactions which may consume about six academic hours (see also Table 2). Because the statistical thermodynamics of equilibrium state is already included in physical chemistry, we give a comprehensively introduction of various ensembles, deriving kinetics of a macroscopic gas system. Starting from the equilibrium theory, we then comprehensively introduce Poincaré's theorem and H-theorem to show the statistical theory of motion. Then, on the basis of quantum mechanics discussed in the last part, statistical theory of reaction rate is quickly given in detail as given in Table 2, such as transition state theory (TST) and molecular dynamics. These topics are often given in physical chemistry or other course. After introducing statistical mechanics, we can show the standard model in chemistry, that is chemistry dynamics and kinetics as the last part.

    As is well known, motion of a microscopic system, such as atom and molecule, must be described by quantum mechanical EOM, namely TDSE. To resolve TDSE of a reaction system, noting that mass of electron is much smaller than that of atom, the Born-Oppenheimer approximation (BOA) is first given (see also Table 1), which needs about three academic hours. Under the BOA, before considering the nuclear dynamics, one has to first solve the electron Schrödinger equation to obtain electron energy of the system with a set of fixing nuclear coordinates. Then, on the basis of a database of extensive electron energy points, the PES can be constructed by numerical regression or interpolation. Having built the PES, we can further solve the nuclear TDSE to obtain the time-dependent wave function of nuclear motion. Finally, various expectations are obtained by the time-dependent wave function and then the flux analyses are used for reactive probability. This scheme of studying chemical

reactions is the so-called "standard model" in theoretical chemistry. We give the teaching content of the standard model in Table 2 as the last part of the present TC course.

However, the numerical implementation of the standard model has an exponential relationship with the degrees of freedom (DOFs) of the reaction system, which is $3N$-6 or $3N$-5 with $N$ number of atoms. This exponential relation greatly limits the application of the standard model for complex poly-atomic reaction. To solve this problem, many theoretical models and computing techniques have developed. The easiest solution is the classical or quasi-classical solution, which is the classical limit of the standard model. According to quasi-classical scheme, the electron Schrödinger equation is no longer solved and thus the PES of large molecules is reduced by model functions, such as force fields, and nuclear dynamics is obtained from Newton EOM. The computational cost of such quasi-classical scheme generally only has a linear relationship with the number of DOFs. Due to this advantage of the quasi-classical scheme, it has become a successful and popular method in simulating poly-atomic molecules. However, the disadvantage of the quasi-classical scheme is that it ignores the quantum characteristics of chemistry reaction, such as zero-point energy, tunneling effect, *etc*. Therefore, the quasi-classical scheme is not *ab initio* scheme, which should be clearly instructed to students.

Finally, due to the contradiction between implementation of standard model and number of DOFs, it is necessary to show how to develop a new model scheme based on quantum mechanics so that the standard model can be applied to macro-molecules. Therefore, the reduction scheme of quantum dynamics for macro-molecules has to be proposed here. To resolve this problem, a reduced dimensional model should be proposed. In the present TC course, as an example of path-integral quantum mechanics, a reduction scheme for cyclic polymer is given just as a final mini-lecture.

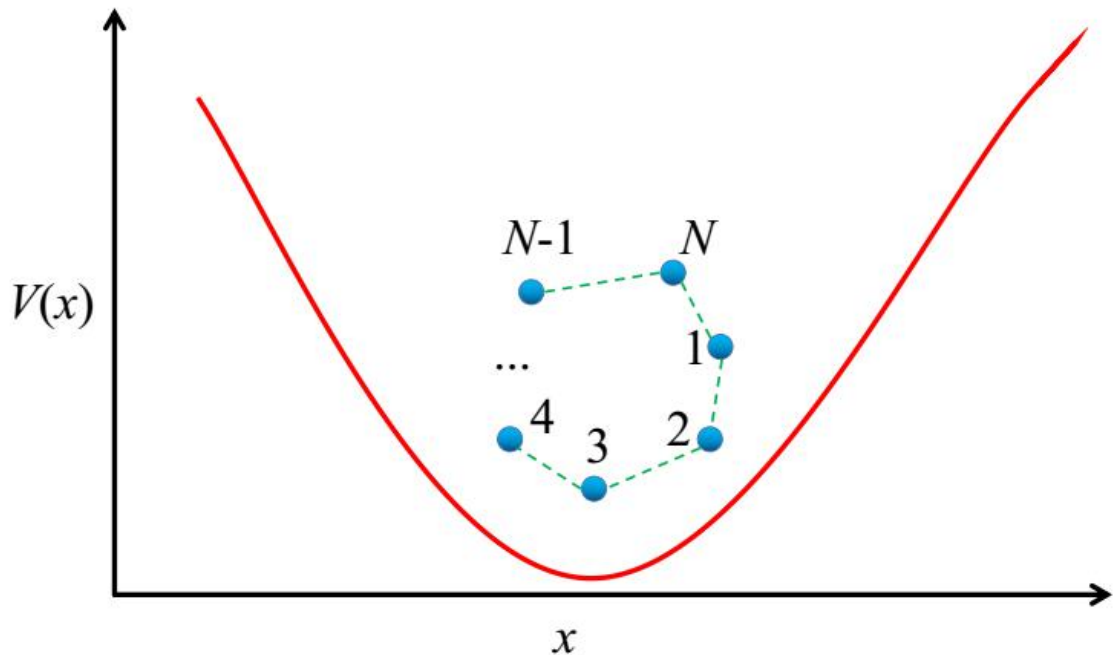

Figure 1. Diagrammatic sketch of the cyclic polymer molecule and its potential energy profile. The red solid line means the potential curve $V(x)$ of monomer of a cyclic polymer.

### An Example for Motion of Polymer

Because of the exponential relation between computational cost and number of DOFs, the standard model for macro-molecule, such as polymer, is hardly implemented. To resolve this problem, a reduced dimensional model has to be used. In the present TC course, as an example of path-integral quantum mechanics, a reduction scheme for cyclic polymer is given. In this scheme, motion of cyclic polymer is reduced to motion of its monomer with the help of transformation of Hamiltonian operators. As shown in Figure 1, let us consider a cyclic polymer with $N$ identical monomers, where $N$ is an enough large number. These $N$ monomers are linked by identical spring. For simple, we assume that the whole molecule lives in a homogeneous chemical environment. Now, let us prove that the motion of this cyclic polymer is equivalent to the motion of all of the monomers connecting with each other by springs. To this end, path-integral quantum mechanics will be used, which is actually a kind of field-theory method for polymer.

First, let us consider the probability amplitude of a particle moving from $x$ to $x'$ in time interval $t$, while $t$ and $x-x'$ are further divided into $\varepsilon$ and $P$ parts, respectively, that is $P\varepsilon = t$. On the basis of path-integral quantum mechanics, the propagator of this particle is[26,27]

$$K(x,x',t) = \langle x|\exp(-iHt/\hbar)|x'\rangle = \lim_{\substack{P\to\infty \\ \varepsilon\to 0^+}} \left(\frac{m}{2i\pi\varepsilon\hbar}\right)^{\frac{P}{2}} \int_x^{x'}\exp\left(\frac{iS}{\hbar}\right)dx_1\cdots dx_P = \int_x^{x'}\exp\left(\frac{iS}{\hbar}\right)\wp x, \quad (1)$$

representing probability of the system moving from $x$ to $x'$ in time interval $t$, where $S$ is action of the particle. According to equation (1), auto-correlation function has the form

$$C(t) = \iint \langle\Psi_0|x\rangle K(x,x',t)\langle x|\Psi_0\rangle dx dx', \quad (2)$$

where $|\Psi_0\rangle$ is the initial state. From auto-correlation function in equation (2), one can obtain dynamics of the particle.

Next, setting $\beta = it/\hbar$ and considering it as temperature,[12,27] equation (1) can be rewritten as

$$\langle x|\exp(-\beta H)|x'\rangle = \int_x^{x'}\exp\left(\frac{S}{\hbar}\right)\wp x, \quad (3)$$

which is the matrix elements of Boltzmann factor $\exp(-\beta H)$. Noting that the Hamiltonian operator of a particle is $H = m\dot{x}^2/2 + V(x)$, the partition function can be written as

$$Z(\beta) = \left(\frac{mP}{2\pi\beta\hbar^2}\right)^{\frac{P}{2}} \int \exp\left\{-\sum_{i=1}^{P}\left[\frac{mP}{2\beta\hbar^2}(x_{i+1}-x_i)^2 + \frac{\beta}{P}V(x_i)\right]\right\} dx_1\cdots dx_P = \left(\frac{mP}{2\pi\beta\hbar^2}\right)^{\frac{P}{2}} \int \exp[-\beta U_{\text{eff}}(x_1,x_2;\cdots,x_P)]\, dx_1\cdots dx_P$$

(4)

where $U_{\text{eff}}(x_1, x_2, \cdots, x_P)$ is effect potential, namely

$$U_{\text{eff}}(x_1, x_2, \cdots, x_P) = \sum_{i=1}^{P}\left[\frac{mP}{2\beta^2\hbar^2}(x_{i+1}-x_i)^2 + \frac{1}{P}V(x_i)\right] = \sum_{i=1}^{P}\left[\frac{m\omega_P^2}{2}(x_{i+1}-x_i)^2 + \frac{1}{P}V(x_i)\right]. \quad (5)$$

We extend equation (4) from the configuration space to the phase space, and hence obtain the form of the partition function on phase space, that is,

$$Z(\beta) = \left(\frac{mP}{2\pi\beta\hbar^2}\right)^{\frac{P}{2}} \int \exp\left[-\beta U_{\text{eff}}(x_1, x_2, \cdots, x_P)\right] dx_1 \cdots dx_P$$

$$\propto \int \exp\left[-\beta \sum_{i=1}^{P} \frac{p_i^2}{2m} - \beta U_{\text{eff}}(x_1, x_2, \cdots, x_P)\right] dx_1 \cdots dx_P dp_1 \cdots dp_P$$

$$= \int \exp\left(-\beta H_{\text{polymer}}\right) dx_1 \cdots dx_P dp_1 \cdots dp_P \qquad (6)$$

where $H_{\text{polymer}}$ is so-called effect Hamiltonian operator of the polymer and, from equations (5) and (6), $H_{\text{polymer}}$ should have the form

$$H_{\text{polymer}} = \sum_{i=1}^{P} \left[\frac{p_i^2}{2m} + \frac{1}{2}m\omega_P^2(x_{i+1}-x_i)^2 + \frac{1}{P}V(x_i)\right]. \qquad (7)$$

In equation (7), the first term represents the total kinetic energy of monomers. The second term is the sum of all interaction between two adjacent monomers. The third term is the total potential energy. Therefore, equation (7) is exactly the Hamiltonian of the cyclic periodic molecule as shown in Figure 1. Comparing equation (4) with (6), the partition functions of a single particle and a cyclic polymer are equal to each other. Therefore, one can conclude that movement of a cyclic polymer is equivalent to the movement of the monomer molecules that are connected with springs. In the present TC course, the above content on cyclic polymer is discussed just after showing the path-integral quantum mechanics as an example. In actual teaching, some mathematical skills might also be mentioned.

### Examples for Experiences

As is well proven, appropriate experiences work well to effectively improve students' learning level. Given in Table 3 are examples of experiences in the present TC course. These experiences need about eight academic hours to solve.

First of all, in the module of classical dynamics, we provide two experiences with H + $H_2$ and H + $CH_4$ as goal reactions. As shown in Table 3, in the first question students are asked to analyze the force of each atom during reaction, which needs students use their knowledge from college physics that taught force analysis. Obviously, this is a unsolvable problem from the viewpoint of mechanics, which can be expected and concluded by students. However, on the basis of our teaching content for Lagrange and Hamilton EOMs, the concept of force is unnecessary here but is given as minus of the gradient of the potential, which is actually concept of force in molecular dynamics. Then, having understood the elemental concept of potential *V* and kinetics energy *T*, the second question for EOMs is very easy to give answer.

Second, in the module of quantum dynamics, we also take H + $H_2$ and H + $CH_4$ as goal reactions and ask students to write Schrödinger equation since the potential *V* and kinetics energy *T* has been given in the classical-dynamics module. This is an easy task since student just needs to understand what factors are *T* and *V* related to. Then, students are asked to show electronic motion in these reactions from the above Schrödinger equation. Students in chemistry often understand the electron and atom transfer processes in these simple reactions. From this question, however, students need to understand how to write electronic Schrödinger equation, where the potential is Coulomb potential. Of course, students may still have no idea on how to resolve it, but they should understand the important of the Schrödinger equation. Finally, students are asked to give Schrödinger equation for reaction, which is related with nuclear coordinates. Students should understand that, by resolving this equation one can obtain wave function of the system and then reactive probability. Comparing the electronic and nuclear Schrödinger equations from questions iv and v, students may find separation of the electronic and nuclear coordinates in the Schrödinger equation is necessary.

Next, in the third module, we asked two questions, including the relation between electron and nuclear and concept of conical intersection. Finally, in the fourth module, two simple questions are asked, and then four questions on chemistry dynamics are asked in the last module. These questions are not difficult even for students in chemistry because these may already be taught in physical chemistry course and other chemistry courses.

**EVALUATIONS**

Obviously, learning TC helps students increasing their understanding for the acquisition of reaction theory.[40] It is also important to train students in understanding the scientific method in chemistry. In the present TC course, we focus on four separate modules, which are given in Tables 1 and 2. Here, we further propose a scheme for evaluating the learning level, including methodology and evaluation resulting. We taught these contents from 2019 to 2020 for 36 students and further evaluated their learning levels through the present scheme. We taught this course twice with the same content and skills.

**Table 4. Evaluative questions for the learning outcome phase. The first column gives the modules of the theoretical chemistry course. The second column gives the evaluative questions. The third and fourth columns give dependent measures of the evaluative questions and their weights, respectively. The fifth column gives the score standard for corresponding questions. The rightmost column gives the overlap of the present questions with physical chemistry or college physics.**

| Modules | Experiences | | | Scoring | Overlap |
|---|---|---|---|---|---|
| | Questions | Dependent Measures | Weights | | |
| Classical Dynamics | I. Classical viewpoint of H + $H_2$ | 1. Description of classical EOM | 30% | know that the classical path is unique | No |
| | | 2. Description of the Lagrangian and Hamiltonian | 40% | know that Newton's law can be derived | No |
| | | 3. Description of difficulties of classical dynamics | 30% | know why path is actually not unique | Yes |
| Quantum Dynamics | II. Quantum viewpoint of H + $H_2$ | 4. Description of quantum EOM | 25% | know Schrödinger equation | Yes |
| | | 5. Description of concept of quantum path | 25% | know uncertainty principle | Yes |
| | | 6. Description of concept of path integral | 20% | know that EOM can be derived | No |
| | | 7. Why system moves as it knows how to solve EOMs | 30% | know the insight of path integral | No |
| Born-Oppenheimer | III. Mechanism of photochemistry | 8. Description of the adiabatic and couplings | 35% | know coupling between two adiabatic states | No |
| | | 9. Description of non-crossing rule | 35% | know the role of group theory in BOA | No |
| | | 10. Description of the mechanism of photochemistry | 30% | know transitional probability between two states | Yes |
| Statistical Ensemble and Kinetics | IV. Rate coefficient of H + $H_2$ | 11. Description of the concept of ensemble | 20% | know why needs ensemble | No |
| | | 12. Description of the Boltzmann distribution of energy | 20% | know rate coefficient from reactive probability | No |
| | | 13. Description of partition function | 20% | know how to obtain partition function | Yes |
| | | 14. Description of Erying equation | 20% | know transition state theory | Yes |
| | | 15. Description of quantum effects of a reaction | 20% | know resonance, interference, *etc.* | No |

### Methodology

In the present work, the learning level was assessed via oral and written tests at the middle-of-term as well as a written exam at the end of term. These tests and exam consisted of four kinds, in total of 15 questions, which are given in Table 4. Comparing Table 3 with Table 4, the experiences (see also Table 3) are different from the evaluation questions (see also Table 4). These evaluation questions are designed to focus on the teaching contents. Most of these questions require essay-style answers or oral representation. In evaluation, we encourage students to send a short essay or oral representation on his/her idea instead of copying from the literature and/or textbook. The generally results are given in Table 5 and illustrated in Figure S1 of Supporting Information. Because objective evaluation of these oral or written tests is difficult, a series of assessment protocol is designed in this work, that is dependent measures for the present evaluation. This series of assessment protocol is also given in Table 4.

Besides the above evaluation questions and dependent measures, we also give in Table 4 the weights of each questions to consider the difficulty of the measures. Moreover, since some contents may be already learned in other courses, we also give the overlap of the assessment protocol with the previous courses. The question whether the present content overlap with the previous courses is also a test question for students, which is designed to ensure a fair assessment. As given in Table 4, the questions pool consisted of four modules, a total of 15 dependent measures. A final 3-point ranking system (*poor*, *fair*, and *good*) resulted, in which a *poor* ranking indicated omitted and/or incorrect answers, and a *good* ranking included comprehensive and excellent answers. Two general measures, namely clarity of writing or oral representation and scientific content of the response, are considered to give the rank. Details of the evaluation method in this work are given in Supporting Information. Finally, students need to give answer on the overlap through *yes* or *no*. These results for each measures are also given in Table 5 and illustrated in Figure S1 of Supporting Information.

**Table 5. Percentage of lowest and highest ratings for each dependent measures, where the categories "Poor", "Fair", and "Good" sum to 100%. The two rightmost columns give students' judgments on whether they have already learned the questions before the present course.**

| Modules | Measures | Rating ($n$ = 36) | | | Overlap ($n$ = 36) | |
|---|---|---|---|---|---|---|
| | | Poor | Fair | Good | Yes | No |
| Classical Dynamics | 1 | 7 (20%) | 21 (58%) | 8 (22%) | 6 (17%) | 30 (83%) |
| | 2 | 15 (42%) | 19 (52%) | 2 (6%) | 0 (0%) | 36 (100%) |
| | 3 | 11 (30%) | 23 (64%) | 2 (6%) | 0 (0%) | 36 (100%) |
| Quantum Dynamics | 4 | 5 (14%) | 11 (30%) | 20 (56%) | 28 (78%) | 8 (22%) |
| | 5 | 11 (30%) | 18 (50%) | 7 (20%) | 19 (53%) | 17 (47%) |
| | 6 | 14 (38%) | 20 (56%) | 2 (6%) | 0 (0%) | 36 (100%) |
| | 7 | 13 (36%) | 19 (53%) | 4 (11%) | 0 (0%) | 36 (100%) |
| Born-Oppenheimer | 8 | 14 (38%) | 21 (59%) | 1 (3%) | 1 (3%) | 35 (97%) |
| | 9 | 15 (41%) | 20 (56%) | 1 (3%) | 2 (6%) | 34 (94%) |
| | 10 | 4 (11%) | 17 (47%) | 15 (42%) | 28 (78%) | 8 (22%) |
| Statistical Ensemble and Kinetics | 11 | 8 (22%) | 25 (69%) | 3 (9%) | 10 (28%) | 26 (72%) |
| | 12 | 1 (3%) | 18 (50%) | 17 (47%) | 19 (53%) | 17 (47%) |
| | 13 | 2 (6%) | 19 (53%) | 15 (41%) | 28 (78%) | 8 (22%) |
| | 14 | 4 (11%) | 20 (56%) | 12 (33%) | 26 (72%) | 10 (28%) |
| | 15 | 10 (28%) | 22 (62%) | 4 (11%) | 16 (44%) | 20 (56%) |

Results

The primary results of the study are summarized in Table 5 as well as in Figure S1 of Supporting Information. Comparing the first module with the second module, one can find that students achieved a slightly higher percentage of fair and good responses (58%~80%) in classical dynamics than those in quantum dynamics (62%~86%). However, they achieved a slightly smaller percentage of poor responses (14%~38%) for the second module (quantum dynamics) than those for the first module (20%~42%). This might be caused by the fact that the beginning of quantum mechanics has been taught in other courses (such as physical chemistry and college physics) but the analytical classical dynamics is not included in the previous courses. One can also conclude this point from the rightmost two columns of Table 5. For example, about 83% of students response that the question #1 is novel, but 78% and 53% of students think that questions #4 and #5 are novel, respectively.

Furthermore, Table 5 shows rather poor teaching results for the module of BOA, partly because this module with a lot of mathematical derivations is a thoroughly new topic for students in chemistry. But fair and good performance (about 89%) for question #10 (that is, mechanism of photo-chemistry) is much better than those for questions #8 and #9 (59%~62%). This is because photo-chemistry is already known by about 78% of students from other chemistry courses, say organic chemistry, while almost all of students (94%~97%) think that questions #8 and #9 include new information. This results imply that the BOA part is a difficult part for learning and needs further teaching designation for clearly and easily understanding.

The situation of the kinetics module (the last module) changes because this module is acquaintance and/or easy to understand for students in chemistry. Many topics in of this module are involved in general physical chemistry course. As shown in Table 5, more than 72% of students give fair and good responses, which is consistent with the fact that more than half of students think these topics were already taught in other courses. For example, question #14 (namely, description of Eyring equation) is one of important contents in physical chemistry, making 89% of students give a level of fair and good. On the other hand, in the present TC courses, there are still 56% of students giving a level of fair. This is partly because we introduce Eyring equation by canonical ensemble, which is not familiar to students. We believe that teaching the same thing from different viewpoints is conducive to cultivating critical thinking of students. However, further design for teaching contents on quantum molecular dynamics and molecular simulation are needed.

Discussions and Future Plan

In the above section, we discussed evaluation results and shown disadvantages of the present teaching contents. Here we propose further designs on teaching content in the near-coming future. They are composed of the following three points.

First, the module of classical dynamics will be reduced to about five academic hours. In the present teaching content, the classical-dynamics module consumes about seven academic hours (see also Table 1), which takes up too much time to learn. To do this, points II and III in Table 1 will be combined into one point with one academic hour to make classical dynamics be easy to understand. This can largely reduce the learning curve for students in chemistry. Moreover, point IV will be reduced to 0.5 academic hour because derivation of Newton EOM is much easier to understand than other points.

Second, two academic hours will be added into the module of BOA to teach this module in more details and more examples, in particular recently development in this field. And then, point XIV takes about 2 academic hours, while point XV takes about 1.5 academic hours. We will also reduce the academic hour of point XVI to 0.5 academic hours because students are familiar with photo-chemistry mechanism. A new example on the geometry phase of reaction will be added, which will consume about 0.5 academic hours. This is because geometry phase of reaction is a frontier of theoretical chemistry.[8]

Finally, points XXV and XXVI will be combined into one point as "reduction of the mechanism", which will consume about one academic hour. To make students easily understand this, we will take the combustion mechanism as a case study because combustion of a fuel (say hydrogen) is an easy case. Moreover, a new point on "state-to-state reaction" with about one academic hour will be added into the module of dynamics and kinetics. This will take the $H + H_2$ reaction as an example to further learn the methodologies discussed in points XXII, XXIII, and XXIV since these three points are not so familiar with students as shown in Table 1.

**CONCLUSIONS**

In this work, we reform the teaching content of theoretical chemistry, establish a teaching sequence from micro- to macro-system, and comprehensively introduce the theory of chemical reaction to undergraduate students. Aiming at the current status of theoretical chemistry teaching, we focus on solving the "last-mile" problem of relationship between the physics and chemistry courses.

We further propose a series of questions that can provide information regarding students' level of knowledge and understanding. It is found that this information is very helpful. In this work, we also analyzed departmental assessment quizzes given to students at the start point of general physical chemistry. It was designed to test students on core concepts from theoretical chemistry that require the use of theory for reactions. By teaching course for 36 students, it was found that students performed significantly improvement from the theoretical chemistry course. Analysis of each individual question revealed that approximately two-third of the students learn new knowledge and understanding from the course.

Future research should focus on understanding how the theoretical chemistry with mathematics and physics can be effectually learned by the students in chemistry, who are often not trained with lots of mathematical derivation. And then, future research should address areas in which performance improvements are still needed, such as recall of factual information and the ability to argue logically from mathematical derivation to properties of chemical reactions and then to theoretical conclusions.

## ASSOCIATED CONTENT
Supporting Information

The Supporting Information is available on the website. In the Supporting Information, we give numerical illustration for Table 5; details of teaching and learning experiences; criteria for evaluation of the learning level.

## AUTHOR INFORMATION

Corresponding Author
*E-mail: qingyong.meng@nwpu.edu.cn

## ACKNOWLEDGMENTS

The author would like to thank the students who participated in our courses for this study. The financial supports of National Natural Science Foundation of China (Grant No. 21773186), Natural Science Foundation of Shaanxi Province (Grant No. 2019JM-380), Teaching Research Funds of Northwestern Polytechnical University (Grant No. 2020JGY32), Shaanxi Research and Practice Projects for Emerging Engineering Education (Grant No. 20GZ110109), and Hundred-Talent Program of Shaanxi are gratefully acknowledged.

# Theoretical Chemistry Course for Students in Chemistry SUPPORTING INFORMATION

Qingyong Meng*

Department of Chemistry, Northwestern Polytechnical University, West Youyi Road 127, 710072 Xi'an, China

## I. Teaching and Learning Experiences

In the present theoretical chemistry (TC) course, we tried our best to adopt a more student centred approach to teaching. By this I mean, we were focusing on students developing some self directed learning skills and becoming more independent in planning of work. We also tried to work on students developing mathematical and physical thinking skills that are very useful in the future research in the field of chemistry.

In teaching the present TC course, we adopted a problem-oriented approach. This means students approached the topic by posing questions or exploring issues which both of students and teacher would attempt to address by completing a number of activities. These include planning and completing some model construction, reading the textbook or reference literature to support students' learning, and discussing with teacher or peers after lecture.

In the present TC course, some of the key outcomes include: (i) a variety of reaction theories in mathematics, (ii) physical thinking skills for chemistry, (iii) problem solving skills. Students would demonstrate these learning outcomes in attempting to address the issues under test and/or discussion. For example, students need to develop new skills in describing the equation of motion in quantum mechanics (see also the fourth dependent measure in Table 4 of the main text). Students not only need to know the mathematical form of this equation, but also the physical background of this equation. In understanding physical background of this equation, the model of path-integral must be introduced, students would no doubt be faced with unexpected challenges which provide them with opportunities to problem solve. Then, these contents on path integral are discussed in the next teaching lecture (see also the fifth and sixth dependent measures in Table 4 of the main text). In order to be able to appreciate the issue, students will be required to read and critically evaluate information from a variety of sources, such as textbook, literature, and peers. In addition, we also encourage students to develop a number of generic skills including independent learning and thinking skills as well as communication skills.

**II. Assessing Experiences for the Learning Outcomes**

As shown in the main text, the learning level was assessed via oral and written tests at the middle-of-term as well as a written exam at the end of term. These tests and exam consisted of four kinds, in total of 15 questions. In evaluation, we encourage students to send a short essay or oral representation on his/her idea instead of copying from the literature and/or textbook. The generally results are given in Table 5 of the main text and illustrated in Figure S1 of this supporting information. Because objective evaluation of these oral or written tests is difficult, a series of assessment protocol is designed in this work, that is dependent measures for the present evaluation. This series of assessment protocol is also given in Table 4 of the main text. The questions pool consisted of four modules, a total of 15 dependent measures. A final 3-point ranking system (*poor*, *fair*, and *good*) resulted, in which a *poor* ranking indicated omitted and/or incorrect answers, and a *good* ranking included comprehensive and excellent answers.

Besides teaching research for teacher, these dependent measures are really opportunities for students to showcase their learning accomplishments in this course. We seek to apply these dependent measures to assessment, where student's overall understanding and knowledge is determined. By these ranking results, evidence of students' learning level can be concluded. By answering these questions in Table 4 of the main text, students send to us some evidence to indicate what they have learned in the present TC course. An example of evidence of learning for inclusion might be a summary of some papers or chapters in a book they have read, and/or a paragraph highlighting the main points that they have taken from the course. For each dependent measure to assessment, students' evidences of learning might contain the following: (i) one or two pages written summary or oral representation of the answer; (ii) plan of self directed learning (written or oral); (iii) overlap (*yes* or *no*) between the present learning unit and the previous; (iv) reflection on this mode and content of learning. In Table S1 we show criterion for grading the learning level.

Table S1. Criterion for grading the learning level.

| Key Outcomes | Grade | | |
|---|---|---|---|
| | Good | Fair | Poor |
| Content | Content is clearly relevant, focused and informative | Content is mainly relevant and informative | Content is only marginally relevant |
| Knowledge | Shows depth of understanding and demonstrates relevant and accurate detail. Sound conceptual and reasoned approach evident. | Shows a degree of knowledge and awareness of the topic. Knowledge is generally accurate but some ideas are not adequately developed or remain vague. | Shows very little knowledge and awareness of the topic. Some knowledge is accurate but there are several ideas that are vague or incorrect. |
| Thinking | Clearly demonstrates the student's intellectual journey, including problem solving and decision making capabilities. | Occasionally reverts to summarizing the lecture note rather than analyzing the problem. | Basic information in assigned lecture note has not been applied. |
| Representation | Written work and/or oral representation is clear and concise. | Overall written work and/or oral representation is concise and clear. | The written work and/or oral representation is not clear. |
| Reflection | Changes in perceptions clearly documented as evidence of self directed learning. | Changes in perceptions can be documented as evidence of self directed learning. | There is little change in perceptions that can be documented as evidence of self directed learning. |

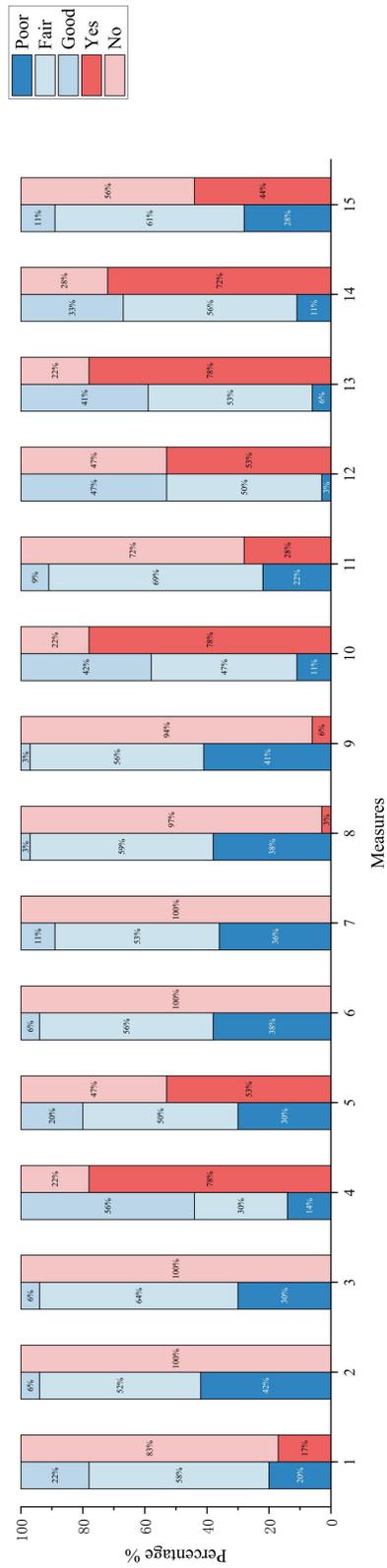

Figure S1. Evaluation data for this activity from the present TC courses, which is also given in Table 5 of the main text.